\documentclass[a4paper]{jpconf}

\pdfoutput=1

\usepackage{graphicx}

\usepackage{hyperref}

\usepackage{amsmath}
\usepackage{amssymb}

\usepackage{algorithm}
\usepackage{algpseudocode}

\usepackage{subcaption}

\DeclareMathOperator*{\rr}{Rejection\,Rate}
\DeclareMathOperator*{\lr}{Loss\,Rate}
\DeclareMathOperator*{\pr}{Pollution\,Rate}

\DeclareMathOperator*{\fn}{False\,Negative}
\DeclareMathOperator*{\fp}{False\,Positive}
\DeclareMathOperator*{\tn}{True\,Negative}
\DeclareMathOperator*{\tp}{True\,Positive}

\begin{document}

\title{Towards automation of data quality system for CERN CMS experiment}
	
\author{M Borisyak$ ^{1, 2}$, F Ratnikov$^{1, 2}$, D Derkach$^{1, 2}$ and A Ustyuzhanin$^{1, 2}$}

\address{$^{1}$ National Research University - Higher School of Economics, 20 Myasnitskaya st., Moscow 101000, Russia}
\address{$^{2}$ Yandex School of Data Analysis, 11/2, Timura Frunze st., Moscow 119021, Russia}

\ead{mborisyak@hse.ru}

\begin{abstract}
Daily operation of a large-scale experiment is a challenging task,
particularly from perspectives of routine monitoring of quality for data being taken.
We describe an approach that uses Machine Learning for the automated system to monitor data quality, which is based on partial use of data qualified manually by detector experts. The system automatically classifies marginal cases: both of good an bad
data, and use human expert decision to classify remaining ``grey area'' cases. 

This study uses collision data collected by the CMS experiment at LHC in 2010.
We demonstrate that proposed workflow is able to automatically process
at least 20\% of samples without noticeable degradation of the result.
\end{abstract}

\section{Introduction}
Data Quality monitoring is a crucial task for every large scale High Energy Physics experiment. The challenge driven by the huge amount of data is the considerable
amount of person power required for monitoring and classification.
An automated system for data quality monitoring can thus save resources needed to keep high quality of collected physics data. 

In these proceedings, we use data collected by the CMS experiment~\cite{chatrchyan2008cms} at LHC in CERN. Currently, CMS data quality is certified by detector experts who base their judgment on the visual inspection of a set of pre-defined distributions. Our system is tested to monitor collision data~\cite{opendataminbias, opendatamu, opendataphoton} collected by the CMS experiment, however the only component of the system dependent from the particular experiment setup is the feature preprocesing step. 

\section{Automated Data Quality System}

A detector measures physical properties of proton collisions products. When a subdetector exposes an abnormal behavior (e.g. part of subdetector becomes unresponsive), it is reflected in measured or reconstructed properties. Data Quality managers rely on some set of statistics and set of rules which describe normal values for these statistics: in case an anomaly occurs some of the statistics should show a considerable deviation from its normal values.
Constructing these statistics requires an exhaustive knowledge of the detector properties and possible anomalies. 

We, in contrast, follow an agnostic approach: instead of constructing a set of dedicated statistics, the system is based on physics properties of collected data, measured or reconstructed. This approach was chosen for a number of reasons. First of all, statistics, similar to those used by experts, can be learned directly from data. This gives a possibility for automated detection of anomalies. Secondly, such system is easily adaptable to different experimental setups including modifications in the detector. Finally, an automated approach and the expert statistics are not mutually exclusive, and injection of expert statistics into the feature set is a good starting point for the improvement of the system.
%
%

The primary goal of the system is to assist Data Quality managers by filtering most obvious cases, both positive and negative. CMS data are aggregated into lumisections each corresponding to approximately 23 seconds of data taking. Lumisection is a data granularity for which the data quality flag is defined in the CMS software stack. 

This task requires classification of all samples into three categories:
\begin{itemize}
	\item definitely anomalous (black zone): decision can be made automatically, samples are marked as anomalous;
	\item definitely good (white zone): decision can be made automatically, samples are marked as good;
	\item ambiguous (gray zone): decision can not be made automatically, human intervention is required.
\end{itemize}

Initially, when no data is available, the system classifies all incoming samples as ambiguous. Samples from gray zone are passed for evaluation to the human experts. Evaluation results are then used for retraining the system. In this way, the systems learns to mimic the human expert.

Formally speaking, the systems objective is minimization of fraction
of data samples passed for the human evaluation. i.e. the rejection rate:
\begin{eqnarray}
	\rr = \frac{\mathrm{Rejected}}{\mathrm{Total}} \to \min \label{eq:rr},
\end{eqnarray}
under constraints:
\begin{eqnarray}
	\lr = \frac{\fn}{\tp + \fn} \leq L_0 \label{eq:lr},\\
	\pr = \frac{\fp}{\fp + \tp} \leq P_0 \label{eq:pr},
\end{eqnarray}
where:
\begin{itemize}
	\item $\mathrm{Rejected}$ --- total quantity of samples rejected by the system,
	\item $\mathrm{Total}$ --- total quantity of processed samples,
	\item $\fn$, $\fp$, $\tn$, $\tp$ --- anomalous data classified
          as good, good data classified as anomalous, correctly
          classified good data and anomalous respectively.
\end{itemize}
Quantities $\rr$, $\lr$ and $\pr$ can also be measured as fractions of
total luminosity rather than number of processed lumisections.

Constants $L_0$ and $P_0$ are to be set in advance and are driven by external requirements to the system. These constants define desired quality of label assignment:
maximal fraction of `lost' lumisections, i.e. good ones classified as anomalous (loss rate); and maximal fraction of anomalous lumisections classified as good ones (pollution rate). In this way, the system is forced to process automatically only the most obvious cases.

\section{Data preprocessing}
\label{sec:data}

As mentioned above, CMS data samples are aggregated into lumisections each corresponding to approximately 23 seconds of data taking.
Thus, a lumisection is a set of events, and each event corresponds to one beam crossing.

In the CMS pipeline, all taken events are split into 'streams' according to some criteria.
In this work, we consider only the following streams:
\begin{itemize}
	\item minimal bias stream: prescaled stream of all events~\cite{opendataminbias};
	\item muon stream~\cite{opendatamu}.
	\item photon stream~\cite{opendataphoton};
\end{itemize}

Reconstructed high level objects in the event are may be divided into 'channels' depending on particles type or registering subsystem:
\begin{itemize}
	\item muons;
	\item photons;
	\item particle flow jets;
	\item calorimeter jets.
\end{itemize}

Every object is characterized by its reconstructed physics properties.
In this work,  the following features are considered:
\begin{itemize}
	\item $p_T$ --- traverse momentum;
	\item $\eta$, $\phi$ --- pseudo-rapidity and angle between the transverse direction and the horizontal plane;
	\item $f_x$, $f_y$, $f_z$ --- coordinates of the reconstructed origin;
	\item $m$ --- reconstructed object mass for composed objects.
\end{itemize}

Since only the whole lumisection can be marked as good or anomalous, aggregation of physical features is performed.

All objects in a given channel in one event are sorted in the
descending order by the momentum and $l = 5$ particles with indices:
$0, \left\lfloor \frac{N}{l} \right\rfloor, \left\lfloor\frac{2 N}{l} \right\rfloor, \dots \left\lfloor\frac{(l - 1) N}{l}\right\rfloor$
are selected to represent the channel for this event.
In this way each event is characterized by fixed number of features.
The last step of the preprocessing is to compute statistics for each
feature for the entire lumisection.
It was found experimentally, that among all considered combinations
the best results are produced if we consider:
\begin{itemize}
	\item mean and standard deviation;
	\item 1, 25, 50, 75, 99 one-sided percentiles.
\end{itemize}

Each lumisection is described by a fixed number of features, each feature having a hierarchical description:
\begin{itemize}
	\item stream: minimum bias, photon or muon;
	\item channel: muons, photons, particle flow jets, calorimiter jets;
	\item quantile by momentum: $\frac{1}{5}, \frac{2}{5}, \dots, 1$;
	\item physics property: $p_T$, $\eta$, $\phi$, $f_x$, $f_y$, $f_z$, $m$;
	\item statistics for the lumisection: 1, 25, 50, 75, 99 percentiles, mean, standard deviation.
\end{itemize}

Additionally, features like the number of events in the lumisection
and three components of the vector sum of  momentum for all particles
in the event, and others were also introduced.

\section{Algorithm}
In order to maximize number of automatically processed samples under constraints \eqref{eq:lr} and \eqref{eq:pr}, a strong classifier is needed.
We consider score function of the chosen classifier as a measure of certainty of the decision. 
Essentially, only two thresholds on classifier score, $\tau_{L}$ and $\tau_{R}$, are of primary interest. These thresholds correspond to minimal and maximal score
of a sample to be automatically classified as `good' or `anomalous', respectively.
We estimate those thresholds on samples, already labeled by human experts.
Independent scores, $\hat{y}$, are obtained by cross-validation procedure introduced in \cite{blum1999beating}, and corresponding estimates for $\lr$ and $\pr$ which are labeled as $\hat{L}_\tau(\hat{y}, y)$ and $\hat{P}_\tau(\hat{y}, y)$ respectively are evaluated for each threshold $\tau$. Thresholds $\tau_{L}$ and $\tau_{P}$ are then selected as respectively maximal and minimal values of $\tau$ for which constraints \eqref{eq:lr} and \eqref{eq:pr} are satisfied.

In order to achieve maximal performance, the system memorizes each decision made by human expert, and the subsequent classifier is trained on all samples available. Note, that the system updates classifier every time when new sample has been processed by a human expert.

Pseudo-code for the system is provided in listing \ref{alg:alg}.

\begin{algorithm}[h]
	\caption{A pseudocode for the automated data quality system.}
	\label{alg:alg}
	\begin{algorithmic}
		\Function{Train}{$X$, $y$, $L_0$, $P_0$}
			\State compute scores $\hat{y}$ by $k$-fold cross-validation as in \cite{blum1999beating}
			\State $\tau_{L} = \max \{ \tau \mid \hat{L}_\tau(\hat{y}, y) \leq L_0 \}$
			\State $\tau_{P} = \min \{ \tau \mid \hat{P}_\tau(\hat{y}, y) \leq P_0 \}$
			\State \Return $\tau_{L}$, $\tau_{P}$, classifier trained on $X, y$	
		\EndFunction
		\State
		\Function{AutomatedDataQuality}{$L_0$, $P_0$}
			\State $\tau_L, \tau_P \gets 0, 1$
			\State $\mathrm{classifier} \gets \frac{1}{2}$
			\State $X_{\mathrm{train}} = \varnothing$
			\State $y_{\mathrm{train}} = \varnothing$
			\For{$i = 0, 1, \dots, N$}
				\State $x_i \gets \text{new sample}$
				\State $\hat{y}_i \gets \mathrm{classifier}(x_i)$
				\If{$\hat{y}_i > \tau_{L}$}
					\State classify $x_i$ as good lumisection
				\ElsIf{$\hat{y}_i < \tau_{P}$}
					\State classify $x_i$ as anomalous lumisection
				\Else
					\State $y_i \gets \text{label from human expert}$
					\State $X \gets (X, x_i)$
					\State $y \gets (y, y_i)$
					\State $\tau_L, \tau_P, \mathrm{classifier} \gets \mathrm{Train}(X, y, L_0, P_0)$
				\EndIf
			\EndFor
		\EndFunction
	\end{algorithmic}
\end{algorithm}

\begin{figure}[h]
	\centering
	\includegraphics[width=0.8\textwidth]{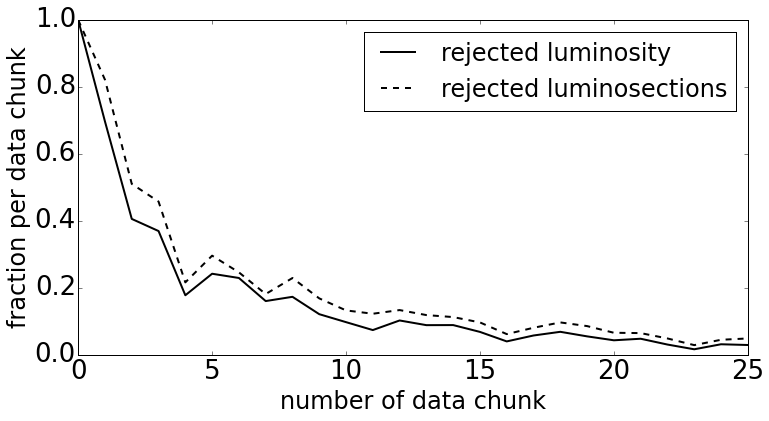}
	\caption{
		Example of learning curves:
		fractions of rejected luminosity and lumisections gradually decrease as 
		classifier gets more labeled data.
	}
	\label{fig:learning}
\end{figure}

It is worth noting that the performance of the system changes over the course of learning,
and it rapidly improves as more data is evaluated and labeled  by the
experts, and thus available for training.
The performance is expected to have some intrinsic limit. An example of empirical learning curves are shown in figure \ref{fig:learning}.

\section{Experiment}
Performance of the system was evaluated on data collected by the CMS
experiment at the LHC in the year 2010 and made available through the CERN OpenData portal (\cite{opendataminbias}, \cite{opendatamu}, \cite{opendataphoton}).
The data was preprocessed as described in section \ref{sec:data}.
The system is implemented with the Gradient Tree Boosting classifier \cite{friedman2001greedy} as
an underlying classifier. 10-fold cross-validation scheme was used to estimate thresholds $\tau_{L}$ and $\tau_{P}$ $^\text{\footnotemark}$.

\footnotetext{The code of the system is available at \url{https://github.com/yandexdataschool/cms-dqm/}.}

\begin{figure}[h]
	\centering
	\begin{subfigure}{0.4\textwidth}
		\includegraphics[width=\textwidth]{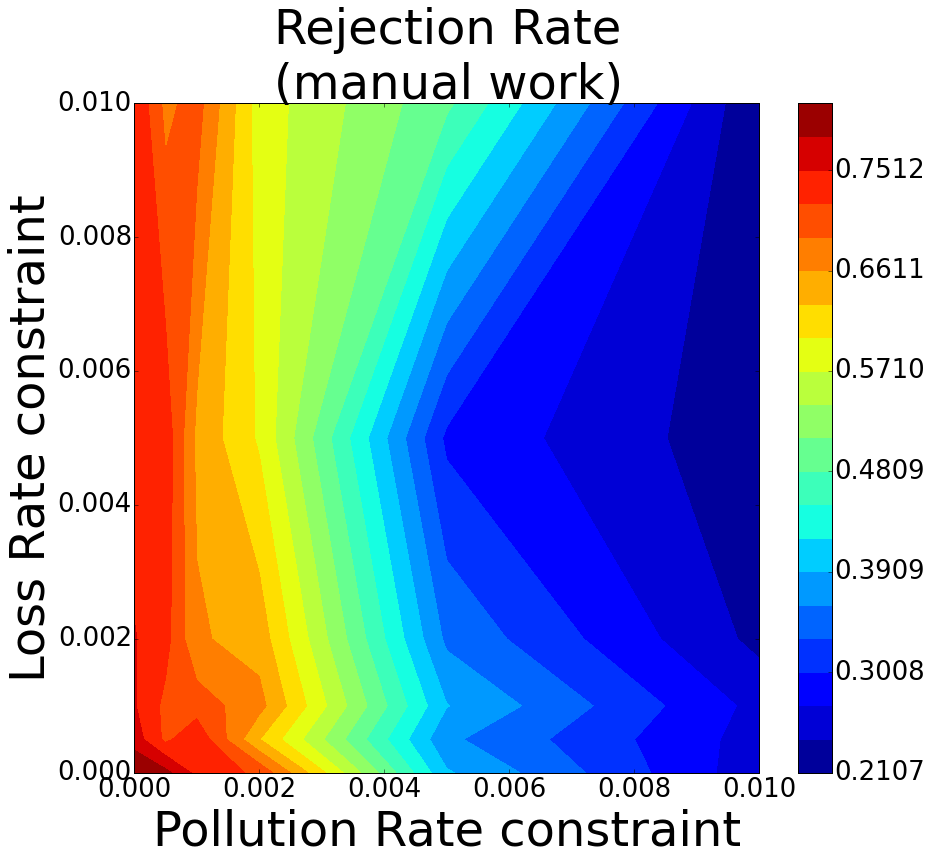}
		\caption{Fraction of rejected samples.}
		\label{fig:results:a}
	\end{subfigure}
	~
	\begin{subfigure}{0.4\textwidth}
		\includegraphics[width=\textwidth]{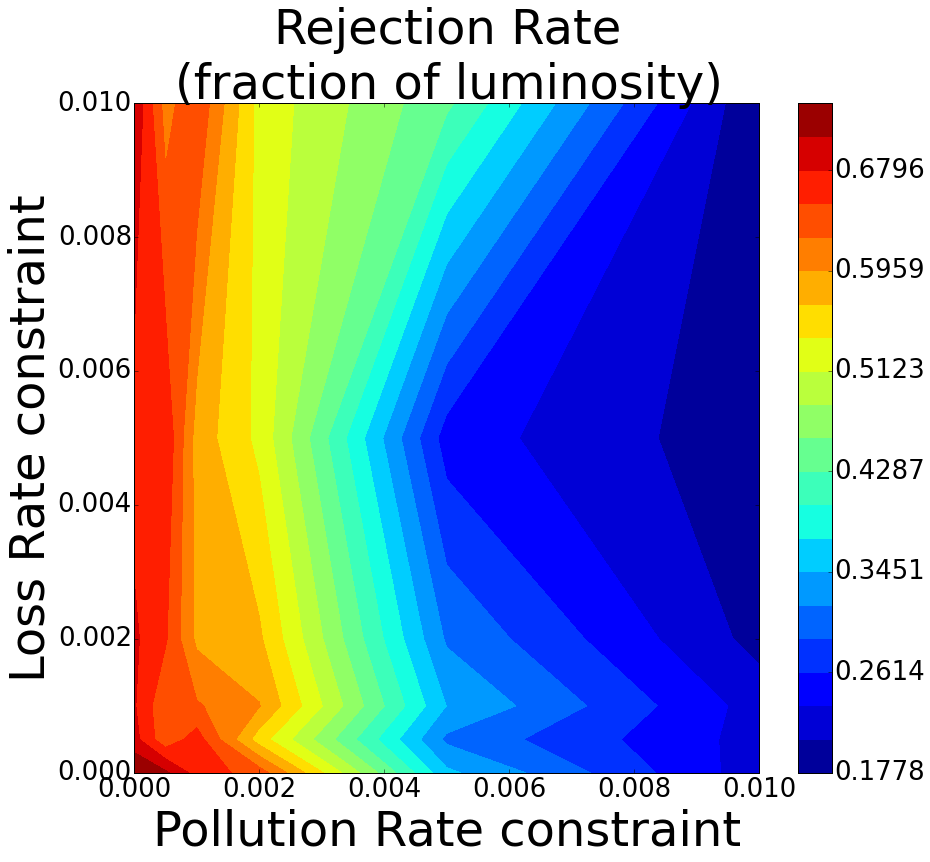}
		\caption{Fraction of rejected luminosity.}
		\label{fig:results:b}
	\end{subfigure}
	\caption{
		Evaluation of the system using CMS experimental data
                published in the CERN open data portal.
		X and Y axis of plots correspond to $\pr$ and $\lr$
                respectively, color corresponds to rejection rate
                expressed as
		\ref{fig:results:a} fraction of rejected samples and \ref{fig:results:b}
		fraction of rejected luminosity.
	}
	\label{fig:results}
\end{figure}

To speed up the evaluation process the data was split randomly into 26 chunks and
the system accepted a whole chunk at once rather than a single sample.
This procedure might result in slight underestimation of system performance, since in this case evaluation does not take into account the data from onr chunk (see~ Alg.~\ref{alg:alg}). However, since a typical time scale of data taking in the CMS experiment is much larger than limitations of this experiment, the number of samples per chunk can be considerably decreased.

The evaluation was performed for constraint on $\pr$ and $\lr$ set to 0,~$1\cdot10^{-3}$,~$2\cdot10^{-3}$,~$5\cdot10^{-3}$,~$10^{-2}$ as fractions of luminosity. Constraint were not violated with the exception of the point $\lr=0, \pr=0$, where violations are present but negligible (less that $10^{-4}$). The system was able to automatically process at least 20\% of samples (which account for 30\% of total luminosity). The rapid growth is observed in both quantities as restrictions become less strict.

\section{Conclusion}
In this work, we described an approach for automated data quality system.
While developed with the CMS experiment in mind, we use an agnostic
approach which allows the straightforward
adaptation of the proposed algorithm to different experimental
setups. We also define the clear strategy for improving performance
using knowledge about detector specifics.
Performance of the system was evaluated on the data collected by the CMS experiment at the LHC in 2010 and made available through the CERN OpenData portal.
Experiments demonstrate that the system is able to automatically
process at least 20\% of samples and 30\% of total luminosity keeping
pollution and loss rates on negligible level, and with more relaxed  restrictions on pollution and loss the performance
of the system significantly increases.

\section*{Acknowledgements}
We wish to thank Maurizio Pierini and Jean-Roch V. Vlimant for their help and support of this study.

\section*{References}
\bibliographystyle{iopart-num}
\bibliography{iopart-num.bib}

\end{document}